\documentclass[12pt]{article}

\usepackage{amsmath}
\usepackage{latexsym}
\newcommand\be{\begin{equation}}
\newcommand\ee{\end{equation}}
\newcommand\bea{\begin{eqnarray}}
\newcommand\eea{\end{eqnarray}}
\newcommand\beas{\begin{eqnarray*}}
\newcommand\eeas{\end{eqnarray*}}

\def\tr{{\rm Tr}}

\def\Xint#1{\mathchoice
{\XXint\displaystyle\textstyle{#1}}%
{\XXint\textstyle\scriptstyle{#1}}%
{\XXint\scriptstyle\scriptscriptstyle{#1}}%
{\XXint\scriptscriptstyle\scriptscriptstyle{#1}}%
\!\int}
\def\XXint#1#2#3{{\setbox0=\hbox{$#1{#2#3}{\int}$ }
\vcenter{\hbox{$#2#3$ }}\kern-.5\wd0}}

\def\dashint{\Xint-}

\begin{document}
\title{How universal is the Wigner distribution?\footnote{WITS-CTP-077}}

\author{\\
Mthokozisi Masuku and Jo\~ao P. Rodrigues\footnote{Email: joao.rodrigues@wits.ac.za} \\
\\
National Institute for Theoretical Physics \\
School of Physics and Centre for Theoretical Physics \\
University of the Witwatersrand, Johannesburg\\
Wits 2050, South Africa 
}

\maketitle

\begin{abstract}
We consider Gaussian ensembles of $m$ $N\times N$ complex matrices. We identify an enhanced symmetry in the system and 
the resultant closed subsector, which is naturally associated with the radial sector of the theory. 
The density of radial eigenvalues is obtained in the large $N$ limit. It is of the Wigner form only for $m=1$. For $m\ge 2$, the new
form of the density is obtained. 
\end{abstract}

\newpage

\section{Introduction}

The Wigner semicircle distribution \cite{Wigner:1955}

$$
f(x) = \frac{2}{\pi R^2} \sqrt{R^2-x^2}, ~ -R < x < R; \qquad f(x)=0,  ~ |x|>R~,
$$

\noindent
describes the density of eigenvalues of a gaussian ensemble of single large hermitean, symmetric or quaternionic matrices \cite{Mehta:2004}, and finds applications in the descripiton of systems in many areas of Physics, from Nuclear Physics to Condensed Matter Physics. 

The Gaussian ensemble of a single complex matrix, or equivalently, of two hermitean matrices, is also described by a Wigner type distribution of 
eigenvalues of a ``radial" variable \cite{Mtho} \cite{Masuku:2009qf}, the definition of which will be made precise in the following.

It is of great interest and importance to investigate if this continues to be a general property of gaussian ensembles of more matrices, and in general, to study the properties of systems with a finite number of matrices, particularly in their large $N$ limit \cite{'t Hooft:1973jz}. 


The reason for this importance includes, for instance, the fact that, as it has been established already some time ago \cite{Eguchi:1982nm}, \cite{Bhanot:1982sh}, \cite{Gross:1982at}, $QCD$ can be reduced to a finite number of matrices with quenched momenta. 

Of more recent interest, the matrix description of $D$ branes \cite{Polchinski:1995mt} has for instance lead to the proposal that the large $N$ limit of the quantum mechanics of the multi matrix description of D$0$ branes provides a definition of $M$ theory \cite{Banks:1996vh}. In the context of the AdS/CFT duality \cite{Maldacena:1997re}, \cite{Gubser:1998bc}, \cite{Witten:1998qj},  
due to supersymmetry and conformal invariance, correlators of supergravity and $1/2$ BPS states reduce to calculation of free matrix model overlaps 
\cite{Lee:1998bxa}, \cite{Corley:2001zk} or consideration of related matrix hamiltonians \cite{Berenstein:2004kk}. 
For stringy states, in the context of the BMN limit \cite{Berenstein:2002jq} and $\cal{N}$ $=4$ SYM, similar considerations apply 
\cite{Constable:2002hw}, \cite{Beisert:2002ff}, \cite{deMelloKoch:2003pv}. 
A plane-wave matrix theory \cite{Kim:2003rza} is related to the $\cal{N}$$=4$ SYM dilatation operator  
\cite{Beisert:2004ry}.

In this communication we will consider $m$ complex $N \times N$ matrices 

$$
                     Z_A \quad A=1,...,m~,
$$

\noindent
or equivalently, an even number $2m$ of  $N \times N$ hermitian matrices and a gaussian ensemble of such matrices:

\be\label{prob}
p[Z_A^{\dagger},Z_A] = \frac{e^{-\frac{w^2}{2} \tr (\sum_{A=1}^m Z_A^{\dagger}Z_A)}}{Z} ~, 
\quad Z = \int [{dZ_A}^{\dagger} {dZ_A}] 
{e^{-\frac{w^2}{2} \tr (\sum_{A=1}^m Z_A^{\dagger}Z_A)}}.
\ee

We will obtain the large $N$ description of the system in terms of the density of eigenvalues of the matrix 

\be\label{MatF}
\sum_{A=1}^m Z_A^{\dagger}Z_A
\ee

This matrix has a very natural interpretation as a matrix valued radial coordinate. 

We will establish that for $m=1$ the density of eigenvalues of the radial matrix is still described by a Wigner distribution, but that this is no longer the case 
for $m\ge 2$ and obtain the form of the new eigenvalue distribution.  

In order to obtain these densities, we will first need to establish a new result, the measure for the probability density (\ref{prob}) in terms of the eigenvalues $\rho_i=r_i^2, ~i=1,...,N$ of (\ref{MatF}):

\bea
\prod_A [{dZ_A}^{\dagger} {dZ_A}] &=&
C_m  \prod_i d\rho_i\rho_i^{m-1} \prod_{i > j} \rho_i^{m-1} \rho_j^{m-1} (\rho_i - \rho_j)^2 \nonumber \\
&=&
D_m  \prod_i dr_i r_i^{2m-1} \prod_{i > j} r_i^{2m-2} r_j^{2m-2} (r_i^2 - r_j^2)^2 \\
&=&  C_m \prod_i d\rho_i\rho_i^{m-1} \Delta_{RM}^2(\rho_i) = D_m \prod_i dr_i r_i^{2m-1} \Delta_{RM}^2(r_i^2) \nonumber ,
\eea

\noindent
where the antisymmetric product

$$
\Delta_{RM}(\rho_i) \equiv \prod_{i > j}  \rho_i^{\frac{m-1}{2}} \rho_j^{\frac{m-1}{2}} (\rho_i - \rho_j) 
$$

\noindent
generalizes the well known Van der Monde determinant $\Delta = \prod_{i > j} (\rho_i - \rho_j)$, and $C_m$ and $D_m$ are numerical constants.

This comunication is organized as follows: In Section $2$ we restate the problem, identify the enlarged symmetries of the gaussiand ensemble and identify a complete
subset of operators invariant under this enlarged symmetry. In Section $3$, based on the remarkable fact that this subset of invariant operators closes under 
Schwinger Dyson equations, the Jacobian of the transformation to these invariant states and to the eigenvalues of (\ref{MatF}) is obtained. In Section $4$, 
the large $N$ density of states on the positive real line is obtained for $m \ge 2$. In this case, the effective potential contains a new logarithmic term which keeps 
the support of the density of eigenvalues strictly positive, i.e., $0 <\rho_- \le \rho  \le \rho_+$. In Section $5$, the density is appropriately extended to the whole real line, allowing for a common description of both the $m=1$ and $m \ge 2$ cases. For $m=1$, a (restricted) Wigner distribution emerges, whereas for $m \ge 2$ the required two cut solution is shown to agree with that of the previous section, when suitably restricted to the positive real line. In Section $6$, these densities are related 
to the density of zeros of certain polynomials. Section $7$ is left for a summary and brief discussion.

\noindent
\section{Gaussian ensemble and symmetries}

\noindent
As stated in the Introduction, we consider a gaussian ensemble of $m$ complex $N\times N$ matrices

$$
(Z_A)_{ij} ~, ~A=1,...,m.
$$

\noindent
The gaussian potential 

$$
    S_g = \frac{w^2}{2} \tr (\sum_A Z_A^{\dagger}Z_A) 
$$

\noindent
is invariant under the $U(N)^{m+1}$ symmetry

\be\label{sym}
              Z_A \to V_A Z_A V^{\dagger}  ~, ~ A=1,...,m,
\ee

\noindent 
with $V_A ~, A=1,...,m$ and $V$ unitary matrices. 
The potential depends only on the eigenvalues of the positive definite, hermitean matrix 
                        
\be\label{mat}
\sum_A Z_A^{\dagger}Z_A ~,
\ee

\noindent
which are denoted

$$
\rho_i = r_i^2 ~,~ i=1,...,N ~,~ \rho_i \ge 0.
$$

In the context of a physical framework where the $m$ complex matrices are naturally associated with $2m$ hermitean matrix valued coordinates, 
as appropriate, for instance, in the description of the dynamics of branes, these eigenvalues have a natural interpretation as the eigenvalues of a matrix valued radial coordinate.   

The purpose of this communication is to obtain the large $N$ distribution of these eigenvalues for the gaussian partition function

$$
Z = \int \prod_A \prod_{ij} {{dZ_A}^\dagger}_{ij} {dZ_A}_{ij} e^{- S_g} 
$$ 

We will do so by ``integrating out" the ``angular matrix valued degrees of freedom or equivalently, by obtaining the jacobian ${\cal{J}}(\rho_i)$ of the change of variables to the ``radial" eigenvalues: 

$$
Z=\int \prod_i {d\rho_i} {\cal{J}}(\rho_i)~e^{- S_g(\rho_i)}
$$

In order to do so, we will consider correlators of operators that are invariant under the symmetry (\ref{sym}), i.e., in the subsector of the theory with this enlarged symmetry. Such invariant operators can be built as the trace (single traces) of powers of the matrix (\ref{mat}), and hence they depend again on the eigenvalues of this matrix only.

A generating function for such opearators is given by  

$$
\Phi_k = \tr e^{ik \sum_A Z_A^{\dagger}Z_A} = \sum_i e^{ik \rho_i}=\sum_i e^{ik r_i^2} ~,
$$

\noindent
or its fourier transform, the density of eigenvalues:

$$
\Phi(\rho)= \int \frac{dk}{2\pi}e^{-ik\rho} \Phi_k = \sum_i \delta(\rho-\rho_i) = \sum_i \delta(\rho-r_i^2).
$$

It turns that these correlators close in this subsector, as it will be shown in the following by use of Schwinger Dyson equations

\noindent
\section{Jacobian}

Schwinger Dyson equations can be obtained from the identity:

\be
 \int \prod_A \prod_{ij} {{dZ_A}^\dagger}_{ij} {dZ_A}_{ij} 
\frac{\partial}{\partial (Z_A)_{ji}} \left(   \frac{\partial \Phi_k }{\partial (Z_A)^{\dagger}_{ij}}  F[\Phi] e^{- S_g} \right) = 0, 
\ee

where $F[\Phi]$ is an arbitrary product of invariant operators. This yields:

\bea\label{SDO}
< \frac{\partial^2 \Phi_k }{\partial (Z_A)^{\dagger}_{ij}{\partial (Z_A)_{ji}}}  F[\Phi]> + 
<\frac{\partial \Phi_k }{\partial (Z_A)^{\dagger}_{ij}} \frac{\partial  F[\Phi]}{\partial (Z_A)_{ji}}> & & \nonumber \\
- < F[\Phi] \frac{\partial \Phi_k }{\partial (Z_A)^{\dagger}_{ij}} \frac{\partial  S_g}{\partial (Z_A)_{ji}}>&=&0.
\eea

We denoted in the above:

$$
<G[\Phi]> \equiv \int \prod_A \prod_{ij} {{dZ_A}^\dagger}_{ij} {dZ_A}_{ij} ~G[\Phi]~e^{- S_g} = \int [d\Phi ] J(\Phi)~G[\Phi]~e^{- S_g}.
$$

Following \cite{Jevicki:1993rr}, we now consider the identity

$$
\int [d\Phi ] \int dk' \frac{\partial}{\partial \Phi_k'} 
\left( \left[\frac{\partial \Phi_k }{\partial (Z_A)^{\dagger}_{ij}}  \frac{\partial \Phi_{k'}}{\partial (Z_A)_{ji}}\right] ~ J(\Phi) ~ F[\Phi]~e^{- S_g}\right)=0.
$$

Then
\bea\label{SDC}
& &\int dk' 
<\frac{\partial}{\partial \Phi_k'}
\left[\frac{\partial \Phi_k }{\partial (Z_A)^{\dagger}_{ij}}  \frac{\partial \Phi_{k'}}{\partial (Z_A)_{ji}}\right] ~ F[\Phi]> \nonumber \\
&+&\int dk'
< \left[\frac{\partial \Phi_k }{\partial (Z_A)^{\dagger}_{ij}}  \frac{\partial \Phi_{k'}}{\partial (Z_A)_{ji}}\right]
 \frac{\partial \ln J(\Phi)}{\partial \Phi_k'} F[\Phi]> \\
&+&
<\frac{\partial \Phi_k }{\partial (Z_A)^{\dagger}_{ij}} \frac{\partial  F[\Phi]}{\partial (Z_A)_{ji}}> - 
< F[\Phi] \frac{\partial \Phi_k }{\partial (Z_A)^{\dagger}_{ij}} \frac{\partial  S_g}{\partial (Z_A)_{ji}}> = 0 \nonumber
\eea

\noindent
where in the last two terms we used the chain rule.

Comparing (\ref{SDO}) with (\ref{SDC}), which are equivalent for arbitrary $F[\Phi]$, it follows that:

\bea\label{JacK}
\int dk'
\left[\frac{\partial \Phi_k }{\partial (Z_A)^{\dagger}_{ij}}  \frac{\partial \Phi_{k'}}{\partial (Z_A)_{ji}}\right]
 \frac{\partial \ln J(\Phi)}{\partial \Phi_k'} & &  \\
+
\int dk' 
\frac{\partial}{\partial \Phi_k'}
\left[\frac{\partial \Phi_k }{\partial (Z_A)^{\dagger}_{ij}}  \frac{\partial \Phi_{k'}}{\partial (Z_A)_{ji}}\right]
&=&
\frac{\partial^2 \Phi_k }{\partial (Z_A)^{\dagger}_{ij}{\partial (Z_A)_{ji}}} \nonumber 
\eea

\noindent
Fourier transforming, and defining 

\bea
\Omega_{\rho\rho'}&=& \int \frac{dk}{2\pi} \int \frac{dk'}{2\pi} e^{-ik\rho} e^{-ik'\rho'} \left[\frac{\partial \Phi_k }{\partial (Z_A)^{\dagger}_{ij}}  \frac{\partial \Phi_{k'}}{\partial (Z_A)_{ji}}\right] \\
w_{\rho} &=& \int \frac{dk}{2\pi} e^{-ik\rho} \frac{\partial^2 \Phi_k }{\partial (Z_A)^{\dagger}_{ij}{\partial (Z_A)_{ji}}}, \nonumber
\eea

\noindent
the differential equation for the jacobian then takes the form

\be\label{JacX}
\int d\rho' \Omega_{\rho\rho'} \frac{\partial \ln J(\Phi)}{\partial \Phi(\rho')} +
\int d\rho' \frac{\partial \Omega_{\rho\rho'}}{\partial \Phi(\rho')} = w_\rho.
\ee

\noindent
$\Omega_{\rho\rho'}$ and $w_{\rho}$ have been obtained in \cite{Masuku:2009qf}:

\bea
\Omega_{\rho\rho'}&=&  
\partial_{\rho}\partial_{\rho'} \left[ \rho \Phi(\rho) \delta (\rho-\rho')    \right]  \\
w_{\rho} &=& -\partial_{\rho} \left[ \rho \Phi(\rho) \left( 2 \dashint \frac{d\rho' \Phi(\rho')}{\rho-\rho'} + \frac{N(m-1)}{\rho}\right)\right]
\eea

\noindent
As a result, the second term in (\ref{JacX}) vanishes and the Jacobian satisfies:

$$
\partial_{\rho} \frac{\partial}{\partial \Phi(\rho)} ~\ln J = 2 \dashint \frac{d\rho' \Phi(\rho')}{\rho-\rho'} + \frac{N(m-1)}{\rho}
$$

This equation was previously obtained in \cite{Masuku:2009qf}, using collective field theory methods \cite{Jevicki:1979mb}.
The solution is 

$$
\ln J = \int d\rho \Phi(\rho) \dashint d\rho' \Phi(\rho) \ln | \rho-\rho' | + N (m-1) \int d\rho \Phi(\rho) \ln {\rho}
$$

\noindent
In terms of the eigenvalues, 

$$
           J  =     \prod_i \rho_i^{m-1} \prod_{i \ne j} \rho_i^{\frac{m-1}{2}} \rho_j^{\frac{m-1}{2}} |\rho_i - \rho_j| =
\prod_i \rho_i^{m-1} \prod_{i > j} \rho_i^{m-1} \rho_j^{m-1} (\rho_i - \rho_j)^2
$$

Since, up to a constant, $\int [d\Phi] \sim \int \prod_i d\rho_i$, we have obtained the result that we sought to establish:

\bea
\int \prod_A \prod_{ij} {{dZ_A}^\dagger}_{ij} {dZ_A}_{ij} e^{- S}  
&=&  
\int \prod_i {d\rho_i} {\cal{J}}(\rho_i)~e^{- S(\rho_i)} =
\int \prod_i {d\rho_i} J(\rho_i)~e^{- S(\rho_i)} \nonumber \\
&=&
C_m \int \prod_i {d\rho_i} \rho_i^{m-1} 
\left[ \prod_{i > j} \rho_i^{m-1} \rho_j^{m-1} (\rho_i - \rho_j)^2 \right]~e^{- S(\rho_i)}, \nonumber
\eea

\noindent
for potentials invariant under (\ref{sym}).

A couple of comments are in order. When $m=1$, the above result reduces to the result first obtained in \cite{Masuku:2009qf}, where an explicit
parametrization of the $2N$ degrees of freedom of two hermitean matrices was obtained, in terms of radial and angular matrix valued coordinates.
There is a classic result \cite{Ginibre:1965zz} \cite{Mehta:2004} parametrizing a single complex matrix in terms of its complex eigenvalues and upper diagonal matrix. This parametrization of degrees of freedom, useful for holomorphic projections, is different from the one considered in this communication. For $m=1$, the existence of a closed hermitean subsector has also been identified in \cite{Kimura:2009ur}. 

Gaussian ensembles of rectangular $M\times N$ matrices have also been discussed in \cite{Cicuta:1986tn}, \cite{Anderson} and \cite{Feinberg:1996qq}. They can be related to the ensembles discussed in this communication when $M=mN$. In this context, the approach followed is equivalent to using the symmetries of the system to set $m-1$ of the $N\times N$ matrices to zero. This corresponds to a "gauge fixed" treatment, as opposed to the gauge invariant approach described in this communication, for which the eigenvalues $r_i$ have a natural identification as a radial coordinate.  

\section{Large $N$ density of the eigenvalues}

Writing 

$$
Z= \int \prod_i {d\rho_i} {\cal{J}}(\rho_i)~e^{- S_g(\rho_i)} = \int \prod_i {d\rho_i} e^{- S_{eff}} ~,
$$

\noindent
we have 

\bea
Seff&=&  \frac{w^2}{2} \int d\rho \Phi(\rho) ~\rho\\
&-& \int d\rho \Phi(\rho) \dashint d\rho' \Phi(\rho') \ln | \rho-\rho' | - N (m-1) \int d\rho \Phi(\rho) \ln {\rho} ~, \nonumber \\ 
& & \int d\rho \Phi(\rho)=N ~. \nonumber
\eea

In order to exhibit explicitly the $N$ dependence, we rescale $\rho \to N \rho$ and $\Phi \to \Phi$, so that 

\bea
Seff&=&  N^2 [ \frac{w^2}{2} \int d\rho \Phi(\rho) ~\rho \\
&-& \int d\rho \Phi(\rho) \dashint d\rho' \Phi(\rho') \ln | \rho-\rho' | - (m-1) \int d\rho \Phi(\rho) \ln {\rho} ] ~, \nonumber \\
& & \int d\rho \Phi(x)=1 \nonumber
\eea

The large $N \to \infty$ configuration is then determined by the stationary condition  ${\partial_\rho} \frac{\partial S_{eff}}{\partial \Phi(\rho)} = 0$, or

\be\label{Stat}
   \dashint \frac{d\rho' \Phi(\rho')}{\rho-\rho'} = \frac{w^2}{4} - \frac{(m-1)}{2\rho}
\ee

We note a major difference between the case of one complex matrix ($m=1$) and the case of more than one complex matrix: for more than 
one complex matrix, an additional logarithmic potential is present, in addition to the standard Van der Monde repulsion amongst the eigenvalues. As a result, the eigenvalues are ``pushed away" from $\rho=0$. 

The $m=1$ solution will be described in the next section, where the range of the density of eigenvalues is extended to the full real line. In this section, we obtain the the density of eigenvalues for $m \ge 2$. 

The solution to the integral equation (\ref{Stat}), generalizing that associated with Penner potentials \cite{Penner:1986}, is obtained using standard methods \cite{Brezin:1977sv}, together with a careful treatment of the $\rho \to 0$ behaviour, following Tan \cite{Tan:1991ay}. 
One introduces the function $G(z)$ in the complex plane

$$
         G(z) = \int_{\rho_-}^{\rho_+} \frac{d\rho' \Phi(\rho')}{z-\rho'}  
$$     

\noindent
where $\Phi(\rho)$ has support only in the interval $[ {\rho_-}, {\rho_+}]$, ${\rho_+} > {\rho_-} > 0$. For large $|z|$, $G(z) \sim \frac{1}{z}$, 
and as $z$ approaches the support of $\Phi(\rho)$,

$$
         G(\rho \pm i \eta) = \dashint_{\rho_-}^{\rho_+} \frac{d\rho' \Phi(\rho')}{\rho-\rho'} \mp i \pi \Phi(\rho) = \frac{w^2}{4} - \frac{(m-1)}{2\rho} \mp i \pi \Phi(\rho).
$$

\noindent
Therefore, this suggests the single cut anzatz 

$$
           G(z) = \frac{w^2}{4} - \frac{(m-1)}{2z} -  \frac{w^2}{4z} \sqrt{(z-\rho_-)(z-\rho_+)}
$$

\noindent
One also requires \cite{Tan:1991ay} that $G(z)$ has no pole as $z\to 0$. These conditions fix:

\be\label{EndP}
              \rho_{\pm} = \frac{2}{w^2}(m+1) \pm \frac{4}{w^2} \sqrt{m}
\ee

\noindent
It follows that the density of eigenvalues is given by

\bea\label{Mtwo}
\Phi(\rho)&=& \frac{w^2}{4 \pi \rho} \sqrt{(\rho_+-\rho)(\rho-\rho_-)}~,  \quad \rho_- \le \rho \le \rho_+ \nonumber \\
&=& \frac{w^2}{4 \pi} \sqrt{(\frac{\rho_+}{\rho}-1)(1 - \frac{\rho_-}{\rho})} \\
&=& \frac{1}{\pi \rho} \sqrt{1-\frac{w^4}{16}(\rho-\frac{2}{w^2} (m+1))^2}~, \nonumber
\eea

\noindent
no longer of the Wigner form.

\section{Symmetric solutions}

In this section, we extend the domain of definition of the density of eigenvalues, allowing us to provide a unified description of the single complex matrix case ($m=1$) and that of more than two complex matrices ($m\ge 2$).

With $\rho=r^2, ~r>0$, define

$$
   2 r \Phi(r^2)  \equiv \phi(r) \equiv \phi(-r)
$$

\noindent
In this way, for an arbitrary function $f(r^2)$

\be\label{asymp}
      \int_{-\infty}^{+\infty} dr f(r^2) \phi(r) = 2 \int_{0}^{+\infty} d\rho f(\rho) \Phi(\rho) .
\ee

\noindent
Returning to (\ref{Stat}), we remark that  ($\rho=r^2$)

\bea
\dashint_0^{\infty} \frac{d\rho' \Phi(\rho')}{\rho-\rho'} &=& \dashint_0^{\infty} \frac{2 r' dr' \Phi({r'}^2)}{r^2-{r'}^2}
 = \frac{1}{2r} \dashint_0^{\infty} dr' \phi({r'}) ( \frac{1}{r-{r'}}+ \frac{1}{r+{r'}}   ) \nonumber \\
&=&
\frac{1}{2r} \dashint_{-\infty}^{\infty} \frac{dr' \phi({r'})}{r-{r'}}
\eea

\noindent
As a result, (\ref{Stat}) is equivalently written as

\be\label{StatR}
  \dashint_{-\infty}^{\infty} \frac{dr' \phi(r')}{r-r'} = \frac{w^2}{2} r - \frac{(m-1)}{r} ~, \qquad
\int_{-\infty}^{+\infty} dr \phi(r) = 2 
\ee

When $m=1$, this integral equation has the well know Wigner distribution as a solution:

$$
\phi(r) = \frac{w^2}{2\pi} \sqrt{\frac{8}{w^2}-r^2}~, \quad       - \frac{\sqrt{8}}{w} \le r \le   \frac{\sqrt{8}}{w}         
$$

\noindent
We can now write

$$
 \Phi(\rho)
           = \frac{w^2}{4 \pi } \sqrt{\frac{8}{w^2 \rho}-1}
~, \quad   0 \le \rho \le   \frac{{8}}{w^2} 
$$

The symmetric solution of (\ref{StatR}) for $m>2$ has been discussed in \cite{Tan:1991ay} \cite{deMelloKoch:1994ir}. It is a two cut solution, with generating functional     

$$
G(z) = \frac{w^2}{2} z - \frac{(m-1)}{z} -  \frac{w^2}{2z} \sqrt{(z^2-{r_-}^2)(z^2-{r_+}^2)}.
$$

The cuts are in the intervals $[ -{r_+}, -{r_-}]$ and $[ {r_-}, -{r_+}]$, with ${r_+}>{r_-}>0$. The asymptotic condition and the absence of a pole at $z\to 0$ fix 

$$
r^2_{\pm} = \frac{2}{w^2}(m+1) \pm \frac{4}{w^2} \sqrt{m} ~,
$$

\noindent
in perfect agreement with (\ref{EndP}). The density is then \cite{deMelloKoch:1994ir}:

$$
\phi(r)=\frac{w^2}{2\pi |r|} \sqrt{({r_+}^2-r^2)(r^2-{r_-}^2)}~, \quad {r_-}^2 \le r^2 \le {r_+}^2~.
$$

\noindent
This agrees with (\ref{Mtwo}), recalling that $ 2 r \Phi(r^2)  \equiv \phi(r) \equiv \phi(-r)$.

\section{Density of eigenvalues and zeros of polynomials}

It turns out that it is possible to relate the densities obtained in the previous section to the density of zeros of certain 
Laguerre or Hermite polynomials \cite{deMelloKoch:1994ir} \cite{Jevicki:1980zp}. 

We use the results of Calogero's work \cite{Calogero:1977qa}, based on the classical results of Stieltjies \cite{Szego}. They show that the zeros of the Laguerre polynomial
$L_{N}^{\alpha} (x)$ satisfy

\be\label{Calo}
  \sum_{j=1,j\ne i }^{N} \frac{1}{x_i-x_j} = \frac{1}{2} \left( 1 - \frac{1+\alpha}{x_i} \right)
\ee
 
In terms of eigenvalues, equation (\ref{Stat}) takes the form:

\be\label{Stateig}
   \sum_{j=1,j\ne i }^{N}  \frac{1}{\rho_i-\rho_j} = \frac{w^2}{4} - \frac{(m-1)}{2\rho_i}
\ee

\noindent
Comparison of these equations shows that the solutions of (\ref{Stateig}) are the zeros of $L_{N}^{m-2} (\frac{w^2}{2}\rho)~, m \ge 2$ .

It has also been established \cite{Calogero:1977qa} \cite{Szego} that the zeros of the Hermite polynomial $H_{2N}(r)$ satisfy

\be\label{CaloH}
  \sum_{j=-N,j\ne i }^{N} \frac{1}{r_i-r_j} = r_i 
\ee

For simplicity, we have chosen an even polynomial. Comparison of the above equation with (\ref{StatR}), considered when $m=1$ and is expressed in terms of eigenvalues, shows that its solutions are the zeros of $H_{2N}(\frac{w}{\sqrt{2}}r)$ \cite{Jevicki:1980zp}.  

The usual relationship between Hermite and Laguerre polynomials is obtained, in the discrete, by noting that, with $x_i>0$, 
and since $x_{-j}= - x_j$, 
the left hand side of (\ref{CaloH}) can be rewritten as

\bea
\sum_{j=-N,j\ne i }^{N} \frac{1}{r_i-r_j} &=&
\sum_{j=1}^{N} \frac{1}{r_i+r_j} + \sum_{j=1,j\ne i }^{N}  \frac{1}{r_i-r_j} \nonumber \\ 
&=&
\frac{1}{2r_i} + \sum_{j=1,j\ne i }^{N} \frac{2r_i}{r_i^2-r_j^2} ~,
\eea

\noindent
and comparing with (\ref{Calo}).









\section{Summary and discussion}

In this communication, we considered gaussian ensembles of $m$ complex $N\times N$ matrices and identified a closed subsector that is naturally 
associated with the radial sector of the theory. In the large $N$ limit, the ensemble is described in terms of the density of radial evenvalues, and these have been obtained: 

$$
\begin{array}{cc}
m=1 
&
m \ge 2
\\
 \Phi(\rho)
 = \frac{w^2}{4 \pi } \sqrt{\frac{8}{w^2 \rho}-1}
&
\quad \Phi(\rho) =\frac{w^2}{4 \pi} \sqrt{(\frac{\rho_+}{\rho}-1)(1 - \frac{\rho_-}{\rho})} 
\\
\quad   0 \le \rho \le   \frac{{8}}{w^2}
&
\quad \rho_- \le \rho \le \rho_+  
\\
&
\rho_{\pm} = \frac{2}{w^2}(m+1) \pm \frac{4}{w^2} \sqrt{m}~.
\end{array}
$$

Extending to the full line with $\Phi(\rho)d\rho=\phi(r) dr~,~\rho=r^2$, the densities take the form:

$$
\begin{array}{cc}
m=1 
&
m \ge 2
\\
\phi(r) = \frac{w^2}{2\pi} \sqrt{\frac{8}{w^2}-r^2}
&
\quad \phi(r)=\frac{w^2}{2\pi |r|} \sqrt{({r_+}^2-r^2)(r^2-{r_-}^2)}
\\
\quad   - \frac{\sqrt{8}}{w} \le r \le   \frac{\sqrt{8}}{w} 
&
\quad {r_-}^2 \le r^2 \le {r_+}^2 
\\
&
r^2_{\pm} = \frac{2}{w^2}(m+1) \pm \frac{4}{w^2} \sqrt{m}~.
\end{array}
$$

\noindent
A (restricted) Wigner distribution is present only for $m=1$. 

The existence of this closed sector is related to an enhanced $U(N)^{m+1}$ symmetry, and the measure in this subsector has been obtained, 
with a result that generalizes the well known single hermitean matrix Van der Monde determinant.

There are several further areas of study that arise naturally from the study presented here. Perhaps of most interest is an investigation of
further systmes, such as Hamiltoniand and/or interacting systems, where this symmetry is present or where the subsector with this symmetry 
may provide physically relevant truncations. It is also of interest to estabish if the results described in this communication extend smoothly to an odd number of matrices.

\section{Acknowledgements}

We would like to thank Robert de Mello Koch for interest in this project and for comments, and G. Cicuta for pointing out to us references \cite{Cicuta:1986tn}, \cite{Anderson} and \cite{Feinberg:1996qq} after submission of this communication's e-print to the Physics archive.

\end{document}